\documentclass[11pt]{article}

\usepackage[margin=1in]{geometry}
\usepackage{lmodern}
\usepackage{microtype}
\usepackage{setspace}
\usepackage{amsmath,amssymb,mathtools,bm}
\usepackage{amsthm}
\numberwithin{equation}{section}
\usepackage{graphicx}
\usepackage{float}
\usepackage{tikz}
\usetikzlibrary{arrows.meta,positioning}
\usepackage{array}
\usepackage{enumitem}
\usepackage{tabularx}
\usepackage{booktabs}
\usepackage[round,authoryear]{natbib}
\usepackage{hyperref}
\hypersetup{
  colorlinks=true,
  linkcolor=black,
  citecolor=black,
  urlcolor=black,
  pdftitle={Mean-Tilted Intervals: Short Tolerance Intervals},
  pdfauthor={Antonio De Leon, Raquel Prado, and Bruno Sanso}
}

\newcommand{\E}{\mathbb{E}}

\newcommand{\ind}{\mathbf{1}}

\newcommand{\TableStyle}{\footnotesize\setlength{\tabcolsep}{4.5pt}\renewcommand{\arraystretch}{1.12}}
\DeclareMathOperator{\argmin}{arg\,min}
\newtheorem{proposition}{Proposition}
\newtheorem{theorem}{Theorem}
\newtheorem{corollary}{Corollary}
\newcounter{algorithm}

\title{\vspace{-0.6cm}Mean-Tilted Intervals:\\
Short Tolerance Intervals}
\author{Antonio De Leon$^{1}$ \and Raquel Prado$^{1}$ \and Bruno Sans\'o$^{1}$}
\date{\small $^{1}$ Department of Statistics, University of California, Santa Cruz\\
\vspace{0.25em}July 2026}

\begin{document}
\maketitle
\vspace{-1.25em}

\begin{abstract}
\noindent
Intervals with the same probability content can have different endpoint
placements and widths. This matters for tolerance inference, where a reported
interval must also satisfy a repeated-sampling content--confidence statement.
We develop mean-tilted intervals (MTIs), a fixed-content family indexed by
retained-mean balance. The zero-tilt member is the mean-preserving interval
(MPI) induced by the residual-product criterion of Pouplin et al.; nonzero
tilts move through admissible contiguous windows, including
distribution-specific central and shortest intervals.

For tolerance inference, we introduce TCSP, a tolerance-calibrated shortest-path
action. TCSP chooses the retained order-statistic count by distribution-free
scan calibration and reports the shortest closed window at that count. This
keeps the certified interval action separate from generalized-posterior endpoint
summaries. We also study a calibrated MTI-ECM comparator that profiles fitted
content and tilt over a prespecified grid and applies an independent
Dirichlet-process content-probability check. In iid simulations at tolerance
confidence \(0.95\), we compare TCSP, MTI-ECM, Young--Mathew interpolation, and
Wilks intervals across feasible content--sample-size cells and eight continuous
distributions. The study emphasizes skewed distributions, where placement
matters most, and excludes cells where the sample range cannot support the
requested two-sided distribution-free statement.
\end{abstract}

\section{Introduction}
\label{sec:introduction}

A prescribed probability content defines a class of possible intervals. Equal-tailed,
shortest-contiguous, and other intervals may contain the same population fraction
while placing their endpoints in different parts of the distribution. Tolerance
intervals add a second requirement: with confidence \(1-\alpha\), the reported
random interval should contain at least content \(c\) of the population
\citep{Wilks1941ToleranceLimits,Wilks1942StatisticalPrediction,Robbins1944DistributionFreeTolerance}.
Thus content, placement, and repeated-sampling confidence are separate parts of
the problem. Fitting the two conditional quantiles at \((1-c)/2\) and
\((1+c)/2\) is appropriate for a central equal-tailed target
\citep{KoenkerBassett1978}. Shortest-interval and mean-preserving questions
require different functionals, so their validation should be aligned with the
interval being estimated \citep{BrehmerGneiting2021Intervals,FisslerEtAl2021SetValued}.

Pouplin et al. introduced a residual-product criterion under the name Relaxed
Quantile Regression \citep{PouplinEtAl2024RQR}. The loss depends on
\((y-\eta_1)(y-\eta_2)\), whose sign records whether \(y\) lies between two raw
roots. The population score equations identify the unrestricted regular target:
among contiguous intervals with content \(c\), the minimizer is the interval
whose retained mean equals the population mean. We call this target the
mean-preserving interval (MPI). The terminology is descriptive; the original
residual-product construction is due to Pouplin et al.

The same score structure gives a placement family. The mean-tilted interval
(MTI) loss keeps the MPI content score and replaces retained-mean equality by a
fixed retained-mean offset \(\delta\). MPI is the \(\delta=0\) member of the
family. Nonzero tilts move the interval within the content-\(c\) class, and the
equal-tailed and shortest-contiguous intervals correspond to
distribution-specific tilts. The tilt is therefore part of the target
specification, or it is selected by a separate calibration rule before empirical
claims are made.

This target geometry leads to a calibrated tolerance action. The shortest
contiguous interval is the minimum-width member of the fixed-content class, but
a sample shortest window must still satisfy a content--confidence requirement.
TCSP denotes the tolerance-calibrated shortest-path action used here: first
calibrate a retained count from the uniform scan statistic, then report the
shortest closed order-statistic window at that count. The scan calibration
supplies the tolerance statement. Any generalized-posterior calculation after
\((c,\delta)\), or after a selected retained count, describes endpoint
uncertainty for a fixed target; the finite-sample tolerance guarantee still
comes from the calibrated action.

A fixed-target MTI-ECM calculation is included to connect the tolerance study to
the MTI generalized-Bayes construction. For each validation cell, a
prespecified adaptive rule profiles fitted content and retained-mean tilt,
keeps candidates that satisfy a direct Dirichlet-process fixed-interval
content-probability check, and applies a width-stability restriction relative
to TCSP. This makes MTI-ECM a calibrated empirical comparator tied to the MTI
loss. The distribution-free scan certificate remains attached to TCSP. A companion
manuscript develops generalized-Bayes endpoint regression, fixed-feature
readouts, and time-varying endpoint models; the present paper keeps only the
parts needed for fixed-content geometry and iid tolerance validation.

The contributions are threefold. First, the residual-product loss is shown to
target MPI, and MTI is shown to generalize MPI by indexing admissible contiguous
fixed-content windows through retained-mean tilt. Second, this geometry is used
to define TCSP, a scan-calibrated minimum-width empirical tolerance action for
iid continuous samples. Third, a prespecified validation study compares TCSP,
the calibrated MTI-ECM comparator, Young--Mathew interpolation, and Wilks
intervals across feasible sample-size/content cells and a distributional panel
chosen to include symmetric, heavy-tailed, and strongly skewed cases.

The formal setting is intentionally bounded. The tolerance statement in this
paper concerns iid continuous univariate samples and the closed order-statistic
window convention used by TCSP. Regression-family and dynamic tolerance
calibration require additional theory. Generalized-posterior endpoint summaries
are loss-based summaries of interval-root functionals; response prediction would
require a separate response-distribution model. Proper scoring rules remain useful when their target matches the
forecast object \citep{GneitingRaftery2007,Gneiting2011,GneitingKatzfuss2014};
the usual interval score targets an equal-tailed interval and is secondary here
\citep{BrehmerGneiting2021Intervals}.

The paper proceeds as follows. Section~\ref{sec:interval-functionals} defines
the fixed-content interval class. Sections~\ref{sec:mpi-loss}
and~\ref{sec:mti-loss} identify the MPI target and the MTI placement family.
Section~\ref{sec:static-theory-statements} records the empirical balance
identity and the fixed-target MTI-ECM bridge used in the validation study.
Section~\ref{sec:tcsp} defines the scan-calibrated tolerance action and its
finite-sample feasibility condition. Section~\ref{sec:evaluation} gives the confirmatory
validation design, and Section~\ref{sec:discussion} concludes.

\section{Fixed-Content Interval Targets}
\label{sec:interval-functionals}

\begin{figure}[H]
\centering
\includegraphics[width=\textwidth]{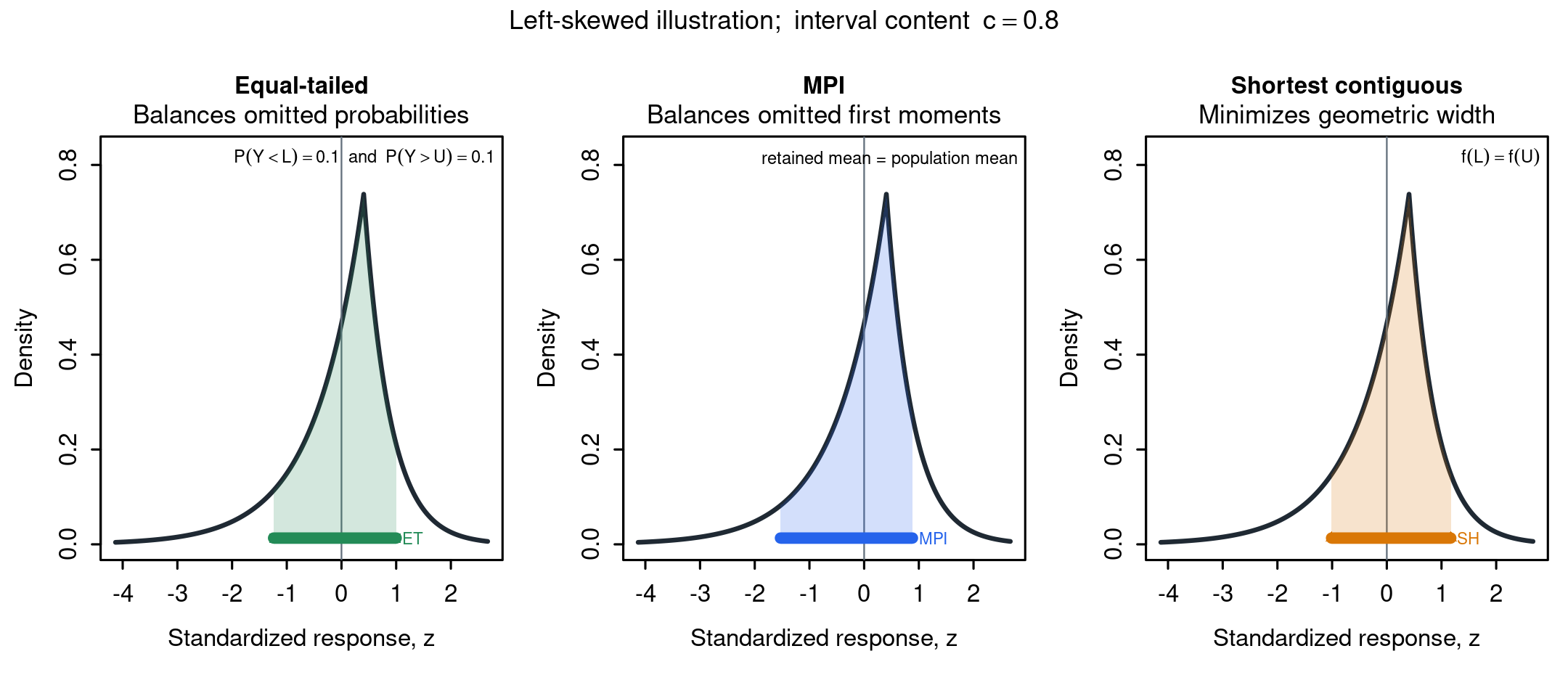}
\caption{\textbf{Three interval-selection principles under left skewness.}
All panels use the same continuous left-skewed population illustration with
content \(c=0.80\), displayed after standardization. Equal-tailed (ET)
intervals balance omitted probabilities, MPI balances omitted first
moments, and shortest-contiguous (SH) intervals minimize geometric width.
These deterministic population calculations compare functionals; they are not
finite-sample or comparative-performance evidence.}
\label{fig:three-interval-principles}
\end{figure}

Let \(Y\) have a nondegenerate continuous distribution with finite mean
\(\mu\) and strictly increasing quantile function \(Q\) on \((0,1)\). Apart
from probability-zero subsets outside the support, every contiguous interval
of content \(c\) can be represented by the probability window
\[
I_c(u)=[Q(u),Q(u+c)],\qquad 0\le u\le1-c.
\]
Define its retained mean and geometric width by
\[
M_c(u)=\frac{1}{c}\int_u^{u+c}Q(v)\,dv,\qquad
W_c(u)=Q(u+c)-Q(u).
\]
Probability content fixes the length of the window on the probability scale,
but not its lower-tail index \(u\). We use ET for equal-tailed, MPI for
mean-preserving, and SH for shortest-contiguous intervals.
Table~\ref{tab:interval-functionals} summarizes these three principal
placement rules.
\begin{table}[h!]
\centering
\caption{\textbf{Three contiguous fixed-content interval functionals.}}
\label{tab:interval-functionals}
\TableStyle
\begin{tabularx}{\textwidth}{@{}>{\raggedright\arraybackslash}p{0.19\textwidth}
  >{\raggedright\arraybackslash}X
  >{\raggedright\arraybackslash}X@{}}
\toprule
Target & Probability-window rule & Placement principle\\
\midrule
Equal-tailed
& \(u=(1-c)/2\)
& Equal omitted tail probabilities\\
MPI
& \(M_c(u)=\mu\)
& Mean-preserving truncation\\
Shortest contiguous
& \(u\in S_{\mathrm{SH}}=\argmin_v W_c(v)\)
& Minimum geometric width\\
\bottomrule
\end{tabularx}
\end{table}
For a regular interior shortest interval under a positive density,
\(f\{Q(u)\}=f\{Q(u+c)\}\). Under multimodality, a highest-density region may
be disconnected, whereas the targets above remain contiguous
\citep{BrehmerGneiting2021Intervals}. Figure~\ref{fig:three-interval-principles}
compares the three principal balance conditions for one strongly asymmetric
distribution.

\section{The Mean-Preserving Interval Target}
\label{sec:mpi-loss}

The residual-product loss becomes a statistical procedure once its target is
identified. This section first defines the pointwise loss, then shows that its
regular unrestricted population solution is the mean-preserving member of the
fixed-content interval class.

Fix a desired interval content \(c\in(0,1)\) and ordered endpoints \(a<b\).
The pointwise MPI loss is
\begin{equation}
\ell_c^{\mathrm{MPI}}(a,b;y)=
\begin{cases}
(1-c)(y-a)(b-y), & a<y<b,\\
c(y-a)(y-b), & y\le a\ \text{or}\ y\ge b.
\end{cases}
\label{eq:mpi-loss-piecewise}
\end{equation}
Both branches are nonnegative and meet continuously at zero when \(y=a\) or
\(y=b\). An observation inside the interval is weighted by \(1-c\), whereas a
miss is weighted by \(c\). The two factors also encode distance from both
endpoints, so the criterion weights membership and endpoint displacement
together. Increasing \(c\) raises the relative cost of a miss; at a regular
population optimum, this weighting
produces the content equation established below. For data
\(y_1,\ldots,y_n\), MPI minimizes
\(\sum_{i=1}^n\ell_c^{\mathrm{MPI}}(a,b;y_i)\), or the corresponding regression
loss when the two endpoints depend on predictors.

The familiar check-loss notation gives an equivalent compact representation.
For a scalar argument \(r\), let
\[
\rho_c(r)=r\{c-\ind(r<0)\}
=
\begin{cases}
(1-c)|r|, & r<0,\\
cr, & r\ge0.
\end{cases}
\]
Now define the product residual
\[
e(a,b;y)=(y-a)(y-b).
\]
Because \(e(a,b;y)<0\) exactly when \(a<y<b\),
\begin{equation}
\ell_c^{\mathrm{MPI}}(a,b;y)=\rho_c\{e(a,b;y)\}.
\label{eq:mpi-product-check}
\end{equation}
This representation also makes clear that the loss is symmetric in its two
roots; the ordering \(a<b\) is imposed only to interpret them as lower and
upper endpoints.
The check loss therefore acts on the scalar product residual, not on
\(y-a\) or \(y-b\) separately. In particular, \(c\) is the desired interval
content; it is neither an assigned response-quantile level for either endpoint
nor the generalized-Bayes learning rate.

The midpoint--half-width parameterization makes the interval geometry
explicit. Write \(a=m-h\) and \(b=m+h\), where \(h>0\). Then
\[
e(a,b;Y)=(Y-m)^2-h^2,
\]
so \(e<0\) means that \(Y\) is within distance \(h\) of \(m\). The half-width
and midpoint are estimated jointly by minimizing the expected loss
\(R_c(a,b)=\E\{\ell_c^{\mathrm{MPI}}(a,b;Y)\}\). Its first-order conditions determine both
the interval content and its placement. The loss defines this population
estimand. When a generalized-Bayes update is used later for a fixed target, its
learning rate controls the strength of the loss update rather than the
population content. The results below concern an unrestricted pointwise
population problem. Restricted regression classes generally recover only
design-weighted projections of these identities.

The endpoint scores have two population consequences: they set the interval
content and make the retained conditional distribution preserve the conditional
mean.

\begin{proposition}[Mean-preserving score characterization]
\label{prop:population-characterization}
Fix \(x\) and let \(a<b\). Suppose
\(\E(Y^2\mid X=x)<\infty\), the conditional distribution has no mass at \(a\)
or \(b\), differentiation may be interchanged with conditional expectation,
and \((a,b)\) is an interior stationary point over an unrestricted pointwise
endpoint class. Then the two population first-order conditions imply
\[
\Pr(a<Y<b\mid X=x)=c,
\]
and
\[
\E\{Y\ind(a<Y<b)\mid X=x\}
=c\,\E(Y\mid X=x).
\]
Equivalently,
\[
\E(Y\mid a<Y<b,X=x)=\E(Y\mid X=x).
\]
\end{proposition}

The population content equation is due to
\citet{PouplinEtAl2024RQR}. The retained-mean equation is a further
consequence of the same first-order system: conditional on \(X=x\), interval
membership has zero covariance with \(Y\). In omitted-tail form, the same
identity balances first-moment mass around the conditional mean:
\[
\E[\{\mu_x-Y\}\ind(Y<a)\mid X=x]
=
\E[\{Y-\mu_x\}\ind(Y>b)\mid X=x],
\qquad \mu_x=\E(Y\mid X=x).
\]
Tail probabilities may differ; MPI balances their first moments around
\(\mu_x\). The retained core and its excluded complement consequently have the
same conditional mean, although the endpoint midpoint may differ from that
mean. The supplement further shows that the retained core is a mean-preserving
contraction of the full distribution in convex order. A finite first moment is
enough for the retained-mean identity once the endpoint scores are valid. The
stronger second-moment assumption ensures that the expected MPI loss
is finite and supports the profiled-risk analysis. This distinction matters
because the loss grows quadratically in an extreme response.

The next theorem converts the score identities into a quantile-window
functional. This is the object that the unrestricted population loss estimates.

\begin{theorem}[Quantile-window identification]
\label{thm:quantile-window}
In addition to the conditions of
Proposition~\ref{prop:population-characterization}, suppose that \(F_x\) is
continuous and strictly increasing on its support, and let \(Q_x=F_x^{-1}\).
Define
\[
M_{c,x}(u)=\frac{1}{c}\int_u^{u+c}Q_x(v)\,dv,
\qquad 0\le u\le1-c.
\]
Then \(M_{c,x}\) is strictly increasing and there is a unique index \(u_c(x)\)
such that \(M_{c,x}\{u_c(x)\}=\mu_x\). The nondegenerate unrestricted MPI
stationary interval is therefore
\[
a_c(x)=Q_x\{u_c(x)\},\qquad
b_c(x)=Q_x\{u_c(x)+c\}.
\]
If \(F_x\) is symmetric about \(\mu_x\), then
\(u_c(x)=(1-c)/2\), and the MPI and equal-tailed intervals coincide.
\end{theorem}

This theorem identifies the unique regular stationary window.
Theorem~\ref{thm:profiled-mean-tilt} below strengthens that statement to
the unique global expected-loss minimizer by profiling over the half-width.
The supplement proves these results and records the restricted score
equations. Under the connected-support and local positive-density conditions
used in Theorem~\ref{thm:profiled-mean-tilt}, the regular unrestricted
population intervals form a nested path. If the conditional quantile function
is continuous at \(p_{\mu,x}=F_x(\mu_x)\) and
\(Q_x(p_{\mu,x})=\mu_x\), the two endpoints converge to \(\mu_x\) as
\(c\downarrow0\). Without this local support condition, the zero-content
limit may straddle a support gap. At fixed positive content, the midpoint's
role remains interval placement, while the conditional mean is imposed through
retained-mean balance. If the
conditional law has endpoint atoms,
the corresponding directional conditions give
\[
\Pr(a<Y<b\mid X=x)\le c\le\Pr(a\le Y\le b\mid X=x),
\]
so exact open-interval coverage requires endpoint continuity. Coincident roots
reduce the loss to \(c(Y-a)^2\) and target the conditional mean rather than a
nontrivial interval. Exact retained-mean equality is not asserted at endpoint
atoms without a separate joint subgradient and mass-allocation analysis.

MPI therefore targets an interval directly and can accommodate asymmetric
conditional distributions without assigning endpoint quantile levels in
advance. The characterization is pointwise and unrestricted; it holds only
under the stated conditions. Linear and other restricted root classes instead
satisfy design-weighted score equations and need not attain conditional
coverage or mean preservation at every covariate value. Nor does the
finite-sample unbiasedness argument for fixed endpoints apply automatically to
fitted endpoints: the endpoints depend on the data, and the empirical
objective is nonsmooth when an observation coincides with a root. Finite-sample
or asymptotic guarantees require a separate empirical-process argument.

\section{Mean-Tilted Intervals}
\label{sec:mti-loss}

The MTI family is the fixed-tilt generalization of MPI. It preserves the MPI
content score and replaces the zero retained-mean-balance equation by a fixed
retained-mean tilt \(\delta\), measured in response units. To move through the
content-\(c\) family without changing the coverage score, define
\[
\ell_{c,\delta}^{\mathrm{MTI}}(m,h;y)
=
\rho_c\{(y-m)^2-h^2\}-2c\delta(m-y).
\]
Equivalently,
\[
\ell_{c,\delta}^{\mathrm{MTI}}(a,b;y)
=
\rho_c\{(y-a)(y-b)\}-c\delta(a+b-2y).
\]
The second form shows that the loss remains invariant to exchanging the two
roots. MPI is exactly the \(\delta=0\) member of the MTI family.

This generalization changes placement but not content. The proposition records
that separation explicitly.

\begin{proposition}[Fixed-content mean tilt]
\label{prop:mean-tilt}
Under the regularity conditions of
Proposition~\ref{prop:population-characterization}, a nondegenerate interior
stationary point of the mean-tilted population risk satisfies
\[
\Pr(a_\delta<Y<b_\delta)=c,\qquad
\E(Y\mid a_\delta<Y<b_\delta)=\mu+\delta.
\]
If \(F\) is continuous and strictly increasing on its support, then for every
\(\delta\) strictly between \(M_c(0)-\mu\) and
\(M_c(1-c)-\mu\), its unique interior quantile-window index solves
\[
M_c(u_\delta)=\mu+\delta.
\]
\end{proposition}

The score equations identify stationary points. Global identification follows
by profiling over the unique content-\(c\) radius at each midpoint and
differentiating the profiled risk; the supplement gives the radius and
envelope-derivative argument.

\begin{theorem}[Global identification of the MTI target]
\label{thm:profiled-mean-tilt}
Suppose that the support closure is an interval (possibly with infinite
endpoints), \(F\) is absolutely continuous and strictly increasing between its
support endpoints, its density is positive and continuous on the support
interior, and \(\E(Y^2)<\infty\). Extend \(F\) by zero and one beyond finite
support endpoints. For every tilt satisfying
\[
M_c(0)-\mu<\delta<M_c(1-c)-\mu,
\]
the expected MTI loss has a unique global minimizer over finite ordered
roots. Its probability-window index \(u_\delta\in(0,1-c)\) is the unique
solution of
\[
M_c(u_\delta)=\mu+\delta.
\]
Nondegeneracy gives \(M_c(0)<\mu<M_c(1-c)\), so MPI is the interior
special case \(\delta=0\).
\end{theorem}

Finite second moments make both endpoint tilt values finite, even when a
support endpoint is infinite. Boundary tilts correspond to one-sided
probability-window limits. Infinite support endpoints lead to semi-infinite
limiting intervals, and finite endpoints require a support constraint to choose
a unique inactive-root representation. Outside the closed admissible tilt range,
the unrestricted risk is unbounded below along the corresponding escaping-root
direction. A proper Gaussian prior can nevertheless yield a proper
finite-sample generalized posterior.

The recovery tilts translate familiar placement rules into the MTI coordinate.
There are three distinct uses of the tilt notation. A fixed target tilt is
chosen before generalized-posterior computation; a population recovery tilt
identifies which \(\delta\) reproduces a familiar interval functional under a
known distribution; an empirical plug-in tilt estimates such a value through a
separate target-selection rule.

\begin{corollary}[Recovery tilts for familiar interval functionals]
\label{cor:recovery-tilts}
MPI is recovered by \(\delta_{\mathrm{MPI}}=0\). Define
\[
S_{\mathrm{SH}}
=
\argmin_{0\le u\le1-c}W_c(u),\qquad
\Delta_{\mathrm{SH}}
=
\{M_c(u)-\mu:u\in S_{\mathrm{SH}}\}.
\]
The population recovery tilt for the equal-tailed interval is
\[
\delta_{\mathrm{ET}}=M_c\{(1-c)/2\}-\mu.
\]
Each \(u\in S_{\mathrm{SH}}\) has recovery tilt
\(\delta_{\mathrm{SH}}(u)=M_c(u)-\mu\). Singular notation is used only when the
shortest-contiguous window is unique or one minimizer has been selected.
\end{corollary}

\begin{figure}[!htbp]
\centering
\includegraphics[width=\textwidth]{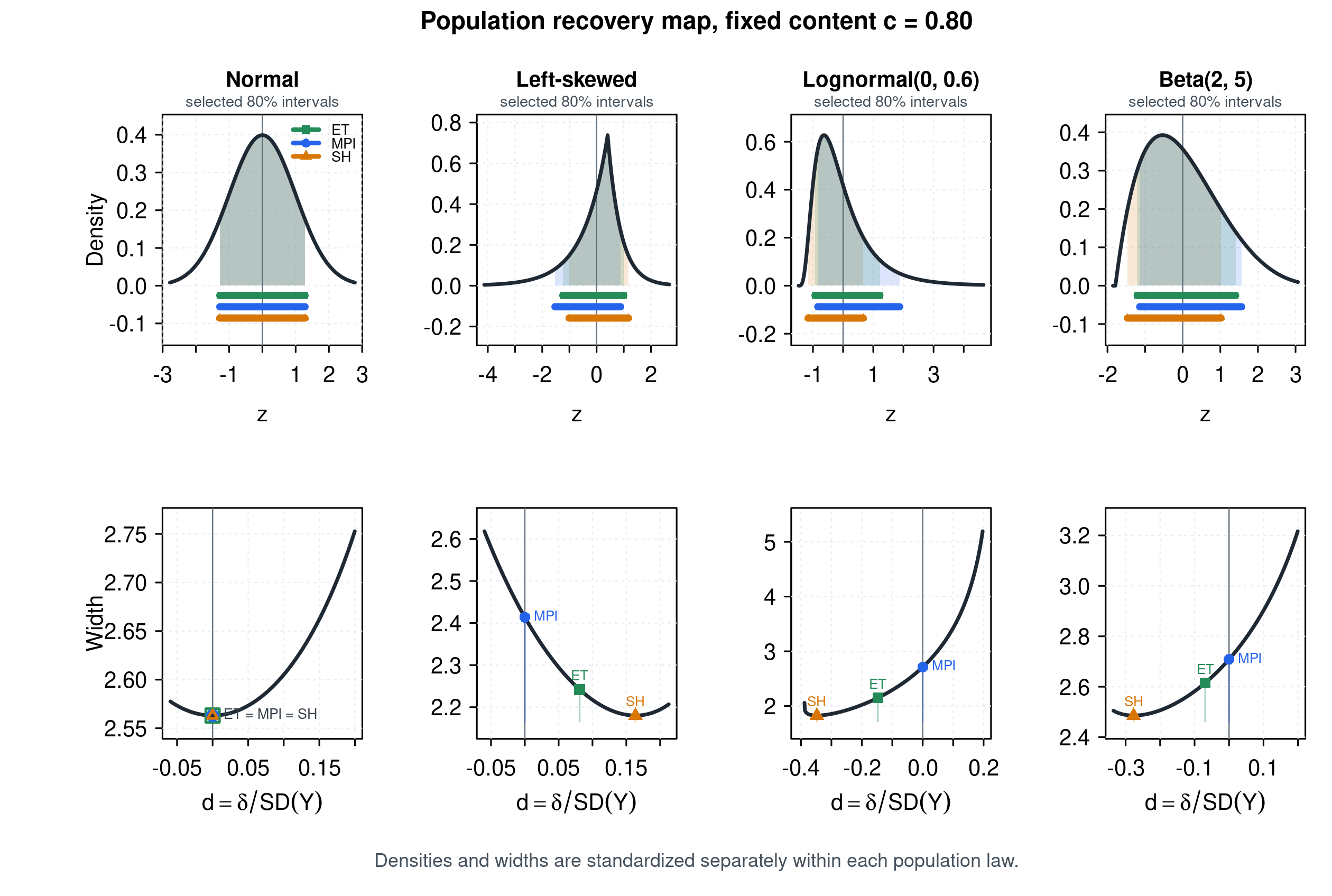}
\caption{\textbf{Population recovery tilts across symmetric and skewed
distributions.}
Every interval has content \(c=0.80\). Targets are computed on each raw
population law and then displayed after mean/standard-deviation
standardization. The top row shows standardized densities with equal-tailed
  (ET), MPI, and shortest-contiguous (SH) intervals; the bottom row
  shows the standardized width profile along the MTI family. In the
Normal reference case the three targets coincide at
\(\delta/\operatorname{SD}(Y)=0\). Under skewness, ET and SH generally require
  nonzero, distribution-specific recovery tilts, while MPI remains the
zero-tilt retained-mean-balance target. The deterministic display is a
population map for external target specification.}
\label{fig:mean-tilt-map}
\end{figure}

At regular interior points where the endpoint quantiles are differentiable
and have positive densities,
\[
\frac{du_\delta}{d\delta}=\frac{c}{W_c(u_\delta)},\qquad
\frac{dW_c(u_\delta)}{d\delta}
=\frac{c\,W_c'(u_\delta)}{W_c(u_\delta)}.
\]
Thus an interior width minimum along the tilt path satisfies the usual
equal-endpoint-density condition.
Figure~\ref{fig:mean-tilt-map} visualizes the same recovery map across
symmetric and skewed population laws before adding any finite-sample
approximation layer.

\begin{figure}[H]
\centering
\includegraphics[width=\textwidth]{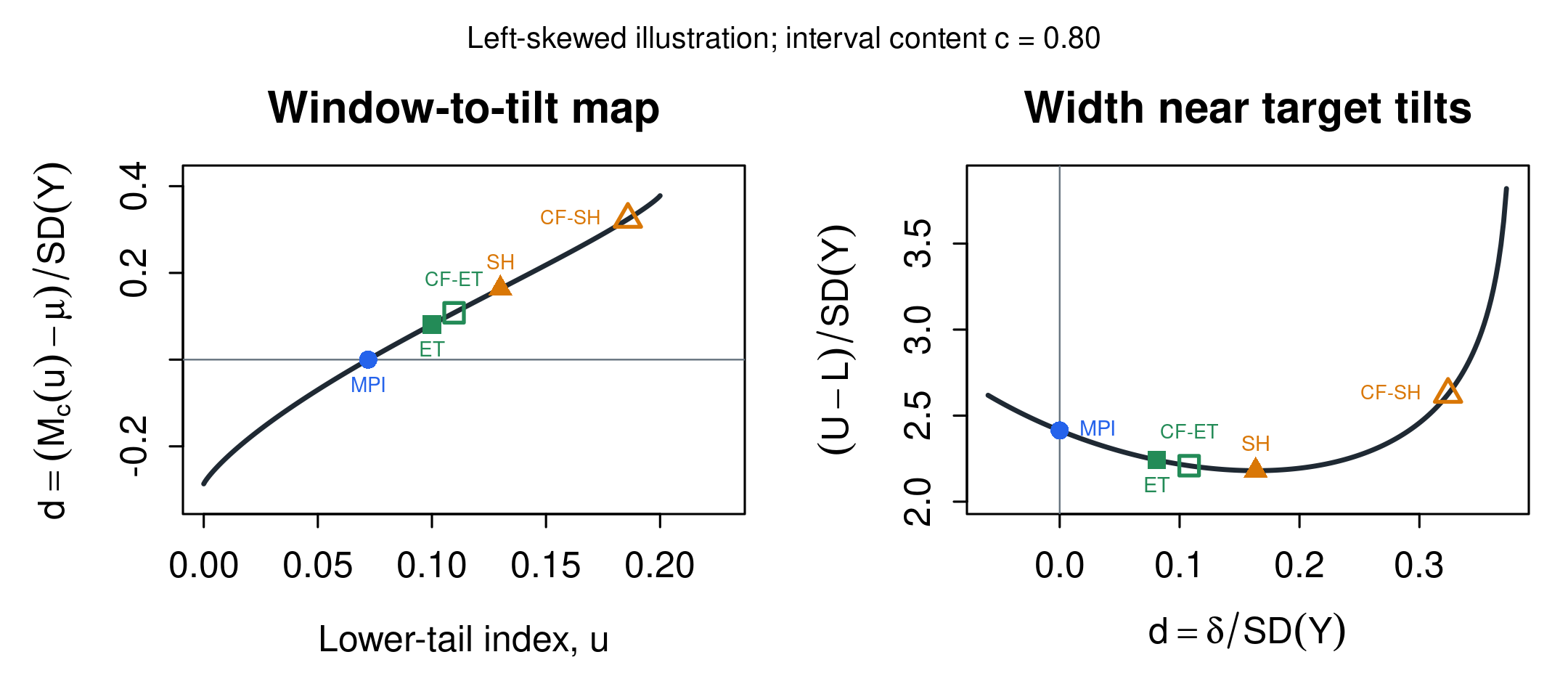}
\caption{\textbf{Cornish--Fisher recovery-tilt approximation.}
For content \(c=0.80\), the solid curves are the population window-to-tilt
map and width profile from the left-skewed illustration. Filled markers are
the exact population MPI, ET, and SH tilts. Open markers are the first-order
Cornish--Fisher approximations computed from true population skewness. The display
diagnoses a population approximation for initialization and external
tilt selection. Data-driven tilt estimation, MCMC evidence, and response prediction
are separate empirical tasks.}
\label{fig:mean-tilt-cf-anchors}
\end{figure}

For initialization and external tilt selection, the same identities suggest a
simple first-order approximation. Let \(\gamma_1\) denote standardized
skewness and set
\[
q_c=\Phi^{-1}\{(1+c)/2\},\qquad K_c=q_c\phi(q_c)/c.
\]
A Cornish--Fisher expansion around a standardized Normal law
\citep{CornishFisher1937} gives the
shortest-oriented and equal-tailed standardized tilt approximations
\[
d_{\mathrm{SH}}^{\mathrm{CF}}=-\gamma_1K_c,\qquad
d_{\mathrm{ET}}^{\mathrm{CF}}=\frac{1}{3}d_{\mathrm{SH}}^{\mathrm{CF}},
\qquad d=\delta/\operatorname{SD}(Y).
\]
These formulas are not new interval targets; they are local skewness-based
approximations to the population recovery tilts. Right skewness gives a
negative shortest-oriented tilt, while left skewness gives a positive tilt.
They supply simple, interpretable starting values for tilt exploration, not
automatic shortest-interval estimators. Figure~\ref{fig:mean-tilt-cf-anchors}
places the two approximations on the exact population tilt path using true
population skewness rather than a sample estimate.

\begin{table}[H]
\centering
\caption{\textbf{Population accuracy of first-order Cornish--Fisher recovery-tilt approximations.} Each row uses a known population law, content $c=0.80$, and true skewness $\gamma_1$. Population and CF columns report standardized recovery tilts $d=\delta/\operatorname{SD}(Y)$; $|e|$ is their absolute gap. Gamma arguments are shape and scale, lognormal arguments are meanlog and sdlog, and the exponential law has unit rate. $^{\dagger}$ denotes an SH population value at the lower support boundary. This is a deterministic population calculation, not MCMC, tilt-selection, or response-prediction evidence.}
\label{tab:mean-tilt-cf-mini-study}
\TableStyle
\begin{tabular}{@{}l@{\hspace{0.8em}}rrrrrrr@{}}
\toprule
Distribution & $\gamma_1$ & \multicolumn{3}{c}{Equal-tailed} & \multicolumn{3}{c}{Shortest-contiguous} \\
\cmidrule(lr){3-5} \cmidrule(l){6-8}
& & Population $d$ & CF $d$ & $|e|$ & Population $d$ & CF $d$ & $|e|$ \\
\midrule
Normal & 0.000 & 0.000 & 0.000 & 0.000 & 0.000 & 0.000 & 0.000 \\
Gamma(16, 0.25) & 0.500 & -0.047 & -0.047 & 0.000 & -0.140 & -0.141 & 0.001 \\
Gamma(4, 1) & 1.000 & -0.092 & -0.094 & 0.002 & -0.273 & -0.281 & 0.008 \\
Lognormal(0, 0.5) & 1.750 & -0.128 & -0.164 & 0.036 & -0.323 & -0.492 & 0.169 \\
Exponential$^{\dagger}$ & 2.000 & -0.169 & -0.187 & 0.018 & -0.402 & -0.562 & 0.160 \\
Beta(2, 5) & 0.596 & -0.069 & -0.056 & 0.013 & -0.277 & -0.168 & 0.109 \\
Beta(5, 2) & -0.596 & 0.069 & 0.056 & 0.013 & 0.277 & 0.168 & 0.109 \\
\bottomrule
\end{tabular}
\end{table}

Table~\ref{tab:mean-tilt-cf-mini-study} shows why the approximation should
remain an approximation rather than a target definition. It is exact at symmetry and
close for the near-Normal Gamma laws. Its errors are larger for lognormal and
bounded Beta shapes and for the exponential law, whose shortest interval is a
support-boundary solution. The reflected Beta rows verify the expected sign
reversal but also show that skewness alone does not determine approximation
accuracy. Across these examples the ET approximation is more accurate than the
SH approximation, but that comparison is descriptive rather than a general
ordering.
The numerical summaries therefore report moment shape,
probability-window boundary behavior, and bootstrap sensitivity whenever a
sample plug-in approximation is constructed.

The recovery tilts above are population quantities. A universal tilt does not
recover the same interval functional across distributions, and conditional
recovery generally requires a covariate-dependent \(\delta(x)\). The tilt is
therefore fixed or selected by a separate validation rule; sampling
it as an ordinary parameter would mix distinct interval targets. The target is
equivariant under positive affine changes of response scale when \(\delta\)
and the fixed learning rate are transformed accordingly, but it is not
invariant to arbitrary monotone response transformations. Generalized-Bayes
computations with nonzero tilt require the target content, tilt, learning rate,
and prior class to be specified before the loss update is evaluated. Other
prior, learning-rate, and model classes require separate propriety arguments and
validation. Data-driven tilt selection likewise requires an independent
selection rule before empirical claims are made.

The population score equations do not by themselves establish propriety of a
finite-sample generalized posterior under an arbitrary root prior. With a
fixed learning rate, a finite design, and proper Gaussian root priors, the tilt
is linear in the root parameters and the Gaussian tails make the generalized
posterior proper. This argument covers ridge regression with proper Gaussian
root priors. Heavier-tailed shrinkage priors and dependent root processes
require separate propriety analyses before nonzero tilt is considered.

\section{Empirical Balance and Fixed-Target Computation}
\label{sec:static-theory-statements}

The population theory above concerns unrestricted fixed-content intervals. A
finite sample adds two separate questions: what the fitted interval balances on
the training data, and how a fixed-target MTI calculation is used after a
content and tilt have been chosen.

\paragraph{Training-sample balance.}
Fitted endpoints depend on the observations, and the empirical MPI/MTI loss is
nonsmooth when an observation lies on a root. The finite-sample score identity is
therefore fractional. For intercept-only MTI with fixed \(\delta\), let
\((\widehat a,\widehat b)\), \(\widehat a<\widehat b\), be an interior
subgradient-stationary empirical solution. Then there exist allocations
\(\widehat q_i\in[0,1]\), equal to one for observations strictly inside the
fitted interval, zero for observations strictly outside it, and fractional only
at fitted roots, such that
\[
\sum_{i=1}^n \widehat q_i=nc,
\qquad
\sum_{i=1}^n \widehat q_i y_i
=nc(\overline y+\delta).
\]
Thus the fitted training interval has exact fractional content and exact
fractional retained-mean balance in its empirical estimating equations.
Population content requires a probability argument beyond this score identity.
In the iid intercept-only case, the Dvoretzky--Kiefer--Wolfowitz inequality gives
a conservative high-probability population-content bound; the supplement records
the explicit display.

\paragraph{Fixed-target generalized-Bayes mode.}
For a fixed fitted content \(q\), tilt \(\delta\), and learning rate
\(\omega_{\mathrm R}\), an MTI generalized posterior updates a prior over
interval roots through the cumulative loss,
\[
\Pi_{q,\delta}(d\theta\mid D_n)
\propto
\exp\{-\omega_{\mathrm R}L_{q,\delta,n}(\theta)\}\Pi_0(d\theta).
\]
In the intercept-only tolerance validation, \(\theta=(a,b)\) and ordered
endpoints are obtained by sorting the two raw roots. This distribution is a
loss-based generalized posterior for a fixed interval target. Response
prediction would require a separate response-distribution model. The tilt
\(\delta\) is fixed before computation or selected by the external calibration
rule described below.

The deterministic MTI-ECM calculation used in the validation study is the mode
calculation associated with this fixed-target update. A pseudo-asymmetric-Laplace
normal--exponential augmentation of the product-residual loss gives exact inverse
latent-scale moments. Conditional on these moments, each root has a Gaussian
conditional maximization step; alternating the two root updates gives an ECM
algorithm \citep{DempsterLairdRubin1977EM,MengRubin1993ECM}. The numerical
implementation uses objective backtracking and deterministic multi-start
selection. The supplement gives the full fixed-target equations. Broader Gibbs,
regression, fixed-feature, and time-varying endpoint models are developed in the
companion computation paper.

\paragraph{Direct content-probability check.}
The MTI-ECM comparator also uses a response-distribution content-probability
check during calibration. For a fixed interval \(I\), a direct
Dirichlet-process posterior for
\(F\) gives an exact Beta law for \(F(I)\) conditional on the observed sample
\citep{Ferguson1973DP}. This check is used to select eligible MTI-ECM
candidates in a separate calibration run. It is a model-based calibration device
for the comparator, while the formal distribution-free tolerance certificate in
this paper remains attached to the TCSP scan action.

\section{Calibrated Minimum-Width Tolerance Intervals}
\label{sec:tcsp}

The preceding theory fixes population content and interval placement. A
tolerance interval adds a repeated-sampling requirement. For iid continuous
univariate observations, the target statement is
\[
\Pr_F^n\{C_F(\widehat T_n)\ge c\}\ge 1-\alpha,\qquad
C_F([L,U])=F(U)-F(L),
\]
where \(\widehat T_n\) is a data-dependent interval. Six quantities must be
kept distinct. The required population content is \(c\), and the tolerance
confidence is \(1-\alpha\). The scan calibration returns a retained count
\(k\), so the fitted content used by a fixed-target MTI approximation is
\(q=k/n\). The fixed MTI placement tilt is \(\delta\), and the
generalized-Bayes learning rate is \(\omega_{\mathrm R}\). When a separate
response-distribution model for \(F\) is used, \(1-\gamma\) denotes a posterior
content-probability threshold for eligible intervals. Neither \(\omega_{\mathrm R}\) nor
\(1-\gamma\) determines the tolerance confidence.
In this article, TCSP denotes the tolerance-calibrated shortest-path action:
first calibrate a retained count by the scan statistic, then report the
shortest closed empirical window at that count.

\paragraph{Scan-calibrated empirical action.}
Let \(Y_{(1)}\le\cdots\le Y_{(n)}\). For content \(c\), define the uniform scan
statistic
\[
M_n(c)=\max_{0\le s\le 1-c}\#\{i:U_i\in[s,s+c]\},
\qquad U_i\stackrel{\mathrm{iid}}{\sim}\mathrm{Unif}(0,1).
\]
The calibrated retained count is
\[
k_{n,c,\alpha}
=\min\{k:\Pr(M_n(c)<k)\ge 1-\alpha\}.
\]
This count is fixed before any endpoint model is fit. The formal empirical
action is the formal closed order-statistic window
\[
\widehat T_n=
[Y_{(\widehat j)},Y_{(\widehat j+k-1)}],
\qquad
\widehat j=\min\argmin_{1\le j\le n-k+1}
\{Y_{(j+k-1)}-Y_{(j)}\},
\]
with \(k=k_{n,c,\alpha}\). This convention is closed and contains exactly
\(k\) sample observations under continuity. It differs from the half-open
\((Y_{(j)},Y_{(j+k)}]\) convention sometimes used for order-statistic
tolerance limits, so a result proved for one convention cannot be transferred
without reconciling the indices and endpoint rule.

\paragraph{MTI summaries after selection.}
The selected window also defines a fixed-target MTI approximation for endpoint
uncertainty. Its fitted content and content buffer are
\[
q_{n,c,\alpha}=k_{n,c,\alpha}/n,\qquad
t_{n,c,\alpha}=q_{n,c,\alpha}-c.
\]
The empirical shortest-path tilt is
\[
\widehat\delta_{\mathrm{SH}}(q)
=
\frac1k\sum_{\ell=\widehat j}^{\widehat j+k-1}Y_{(\ell)}
-\frac1n\sum_{i=1}^nY_i.
\]
After \(q\) and \(\widehat\delta_{\mathrm{SH}}\) are fixed, an MTI endpoint
model with parameter block \(\theta\) uses the loss-based update
\[
\pi_{q,\delta,\omega_{\mathrm R}}(\theta\mid D_n)
\propto
\exp\{-\omega_{\mathrm R}L_{q,\delta,n}(\theta)\}\pi_0(\theta).
\]
This generalized posterior quantifies uncertainty about a fixed interval-root
functional after the target has been specified. The formal tolerance action
remains the empirical order-statistic window \(\widehat T_n\). Because
\(\widehat\delta_{\mathrm{SH}}\) is computed from the same sample, these
posterior summaries are conditional on the selected target and do not propagate
shortest-window selection uncertainty. An unconditional interpretation requires
a plug-in stability theorem or a separate target-selection sample.

\paragraph{Supplementary extensions.}
Response-distribution Bayesian shortest-interval calculations and sample-split spacing constructions
are useful extensions, but they answer different evidentiary questions from the
primary TCSP action. A response-distribution model gives posterior
content probabilities under a model for \(F\), whereas TCSP supplies a
distribution-free scan-calibrated empirical action. A split-spacing construction
uses independent target-selection and calibration samples, whereas TCSP uses the full
sample to choose the minimum-width calibrated order-statistic window. These
extensions remain available as separate methodological and proof directions;
they are not part of the main validation comparison.

\paragraph{Numerical certification and proof guarantees.}
Numerical evaluation of \(k_{n,c,\alpha}\) is consequential because it sets the
reported interval. The calculation uses adaptive Monte Carlo
calibration of the uniform scan distribution. At each look, one-sided
Clopper--Pearson lower bounds are Bonferroni-adjusted over feasible retained
counts and adaptive looks, and the first retained count whose certified lower
probability reaches \(1-\alpha\) is selected. The full-range interval
uses its exact order-statistic probability. A more conservative
Massart--DKW empirical-CDF lower band remains available as a conservative
alternative
\citep{Massart1990DKW}. Exact scan recursion remains an open numerical task.
Accordingly, exact critical-count calculation and the finite-sample theorem for
this closed-window action are treated as open proof and certification tasks.

The available theory is deliberately narrower than the computational procedures.
Theorems~\ref{thm:quantile-window}--\ref{thm:profiled-mean-tilt} identify
the fixed-content MTI family and its admissible tilt coordinate. For tolerance
inference, the empirical closed order-statistic window is the formal action;
generalized-posterior summaries for fixed \((q,\delta)\) describe endpoint-root
uncertainty after a target has been specified and do not serve as the
content--confidence certificate. Split exact-spacing calibration has an exact
conditional iid univariate Beta law. The scan count is certified by the
adaptive Clopper--Pearson numerical procedure above, with numerical confidence
distinguished from tolerance confidence. Direct DP and Gaussian DPM
intervals are ordinary
response-distribution Bayesian alternatives, not MTI generalized posteriors.
Exact scan recursion, an action-matched finite-sample proof for the closed
window convention, selected-action asymptotics, posterior-to-action transfer,
and regression-family tolerance calibration remain separate certification
tasks.

\paragraph{Shortest-path continuation.}
The population shortest-window path gives useful continuation geometry for
moving across contents. This geometry helps initialize and check numerical
paths; it is not the scan certificate. Suppose that the density is
differentiable and strictly unimodal,
and that the shortest window
\(I_q=[L_q,U_q]=[Q(u_q),Q(u_q+q)]\) is unique and interior over a range of
\(q\). Equal endpoint densities then satisfy
\[
f(L_q)=f(U_q)=\lambda_q.
\]
Writing \(a_q=f'(L_q)>0\) and \(b_q=f'(U_q)<0\), formal differentiation of
the content and equal-density equations gives
\[
L_q'=\frac{b_q}{\lambda_q(a_q-b_q)}<0,
\qquad
U_q'=\frac{a_q}{\lambda_q(a_q-b_q)}>0.
\]
Thus the first-order shares of added probability assigned to the left and
right boundaries are
\[
p_L(q)=\frac{-b_q}{a_q-b_q},\qquad
p_R(q)=\frac{a_q}{a_q-b_q},\qquad
p_L(q)+p_R(q)=1.
\]
Only under local symmetry are the two shares equal to one half. The width
satisfies
\[
W_{\mathrm{SH}}'(q)=\frac{1}{\lambda_q},
\qquad
W_{\mathrm{SH}}''(q)
=-\frac{a_qb_q}{\lambda_q^3(a_q-b_q)}>0.
\]
Hence additional content becomes increasingly expensive as the endpoints move
into lower-density regions. If
\(m_q=q^{-1}\int_{u_q}^{u_q+q}Q(s)\,ds
=\mu+\delta_{\mathrm{SH}}(q)\), then
\[
\delta_{\mathrm{SH}}'(q)
=\frac{p_L(q)L_q+p_R(q)U_q-m_q}{q}.
\]
These derivative calculations motivate numerical continuation and diagnose
its behavior. Full regularity, boundary, and nonuniqueness arguments are
reserved for a formal shortest-path theorem.

The continuation calculation starts from a globally solved content, predicts
each next window by nested expansion of the previous window, applies local
correction in a neighborhood of the current solution, expands the neighborhood
when the local optimum lands on its boundary, and finally verifies the formal
action by an exhaustive empirical scan. The exhaustive scan, not the
continuation approximation, defines the selected window. The continuation
argument does not cover infeasible calibration
\((k>n)\), ties or atoms that violate the continuous-sample convention,
nonunique shortest windows, response-scale dependence of geometric width, and
unsupported attempts to attach the scan certificate to posterior summaries.

The construction therefore retains three distinct outputs: an empirical scan
tolerance action, a full-distribution Bayesian shortest-interval uncertainty
quantification action, and fixed-target MTI generalized-posterior summaries.
Finite-sample scan
validity requires an action-matched proof and certified critical count;
first-order tolerance or population-reference width statements additionally require
same-sample shortest-window asymptotics. Any transfer from posterior content
probabilities to endpoint coverage, or from fixed-target MTI summaries to
the selected shortest action, requires a separate action-equivalence or
direct-calibration result. Tolerance confidence comes from the scan calibration,
and \(\omega_{\mathrm R}\) remains the generalized-Bayes learning rate.

\section{Validation Design}
\label{sec:evaluation}

Validation is reserved for questions that require repeated sampling or
held-out comparison. Population calculations verify the fixed-content and
mean-tilt characterizations without Monte Carlo error. Reference calculations
compare generalized-posterior draws with quadrature, dense Gaussian full
conditionals, analytic scale updates, and restart equivalence. These
checks provide numerical evidence for the stated computational target on
reference problems. Empirical coverage, width efficiency, and comparator
ranking require repeated-sampling validation.

The main tolerance study is prespecified before inspecting comparative results,
following simulation-study reporting principles
\citep{MorrisWhiteCrowther2019Simulation}. It uses iid continuous samples from
eight centered and scaled simulation distributions. Gaussian and Student \(t_3\)
samples serve as symmetric reference distributions. The asymmetric simulation
distributions are exponential,
asymmetric Laplace with \(\tau=0.10\), a two-piece Normal law with scale ratio
\(1{:}12\), Beta\((1,8)\), Gamma\((0.5,1)\), and log-normal with log-scale
standard deviation \(1.25\). This panel is intentionally skewed toward
asymmetric laws because the theoretical advantage of a shortest calibrated
window is most relevant when equal-tail or full-range behavior is inefficient.
Multimodal cases are excluded because none of the reported methods is designed
to recover disconnected high-density regions.
The primary simulation design fixes tolerance confidence at \(0.95\) and uses 1000
paired replications per simulation distribution and feasible cell.
The feasible cells are \(n=50,c=0.90\), \(n=100,c\in\{0.90,0.95\}\), and
\(n\in\{500,1000\},c\in\{0.90,0.95,0.99\}\).

The small-sample restriction is structural. For a two-sided distribution-free
interval, even the full sample range reaches content \(c\) with confidence
\(0.95\) only when
\[
  1-n(1-c)c^{n-1}-c^n \ge 0.95 .
\]
The corresponding minimum sample sizes are \(46\), \(93\), and \(473\) for
\(c=0.90\), \(0.95\), and \(0.99\), respectively. Thus
\(n=50,c=0.95\) and \(n=100,c=0.99\) are excluded from the confirmatory
comparison because no two-sided distribution-free method can certify the
requested statement in those cells. The \(c=0.99\) rows begin at \(n=500\), the
nearest rounded sample size above the \(n=473\) threshold.

The comparison is deliberately narrow. TCSP is the formal scan-calibrated
shortest empirical action and is therefore the primary tolerance action. The
reported TCSP calculations use the adaptive Clopper--Pearson scan calibration described
above. Its optimality is class-specific: it minimizes width among closed
order-statistic windows satisfying the calibrated retained-count rule.
The MTI-ECM comparator links the validation study to the generalized-Bayes
interval construction. It uses a prespecified adaptive calibration rule indexed
by sample size and requested content. Within each cell, the rule fixes a direct
Dirichlet-process content-probability check from an independent calibration run, profiles
fixed-target MTI-ECM endpoint fits over fitted content and retained-mean tilt,
and keeps the shortest candidate satisfying the content-probability check and a
same-cell width-stability restriction against TCSP. This empirical validation is
separate from the distribution-free scan theorem that supports TCSP.
Young--Mathew interpolation
\citep{YoungMathew2014NonparametricTolerance} and Wilks min--max intervals
\citep{Wilks1941ToleranceLimits,Wilks1942StatisticalPrediction} provide the
classical order-statistic comparators. The primary metric is the
content-attainment rate: the Monte Carlo estimate of
\(\Pr_F^n\{C_F(\widehat T_n)\ge c\}\), where
\(C_F([L,U])=F(U)-F(L)\). For each replication, this content check uses the
known simulation distribution rather than the fraction of sampled observations
inside the produced interval. If a method produced no interval in a replication,
that replication would be counted as nonattainment. Widths are reported as distribution-level empirical
2.5\%--97.5\% ranges, avoiding width averages across distributions with
different scales and tail behavior.

\begin{table}[!htbp]
\centering
\caption{\textbf{Independent-sample tolerance validation.} Results are shown by
feasible sample-size/content cell and method. Each row summarizes the range
across eight simulation distributions at tolerance confidence \(0.95\). All four
reported methods returned an interval in every replication in this run.}
\label{tab:tolerance-validation-primary}
\TableStyle
\begin{tabular}{@{}l@{\hspace{0.75em}}l@{\hspace{0.75em}}r@{}}
\toprule
Cell & Method & Content-attainment range (\%)\\
\midrule
\(n=50,c=0.90\) & TCSP & 95.6--97.7 \\
 & MTI-ECM & 95.8--97.8 \\
 & Young--Mathew & 94.6--97.0 \\
 & Wilks & 95.6--97.7 \\
\addlinespace[0.25em]
\(n=100,c=0.90\) & TCSP & 98.3--99.4 \\
 & MTI-ECM & 98.5--99.7 \\
 & Young--Mathew & 95.2--96.9 \\
 & Wilks & 99.9--100.0 \\
\addlinespace[0.25em]
\(n=100,c=0.95\) & TCSP & 96.2--97.3 \\
 & MTI-ECM & 95.1--97.5 \\
 & Young--Mathew & 95.5--96.8 \\
 & Wilks & 96.2--97.3 \\
\addlinespace[0.25em]
\(n=500,c=0.90\) & TCSP & 96.7--99.0 \\
 & MTI-ECM & 97.3--99.7 \\
 & Young--Mathew & 94.4--96.1 \\
 & Wilks & 100.0 \\
\addlinespace[0.25em]
\(n=500,c=0.95\) & TCSP & 96.2--99.1 \\
 & MTI-ECM & 97.0--99.2 \\
 & Young--Mathew & 94.2--96.6 \\
 & Wilks & 100.0 \\
\addlinespace[0.25em]
\(n=500,c=0.99\) & TCSP & 95.3--96.7 \\
 & MTI-ECM & 95.2--97.4 \\
 & Young--Mathew & 95.1--96.7 \\
 & Wilks & 95.3--96.7 \\
\addlinespace[0.25em]
\(n=1000,c=0.90\) & TCSP & 98.0--99.6 \\
 & MTI-ECM & 97.7--99.5 \\
 & Young--Mathew & 94.2--96.6 \\
 & Wilks & 100.0 \\
\addlinespace[0.25em]
\(n=1000,c=0.95\) & TCSP & 97.6--99.6 \\
 & MTI-ECM & 96.7--99.9 \\
 & Young--Mathew & 95.1--96.3 \\
 & Wilks & 100.0 \\
\addlinespace[0.25em]
\(n=1000,c=0.99\) & TCSP & 97.5--99.5 \\
 & MTI-ECM & 97.9--99.4 \\
 & Young--Mathew & 94.1--97.2 \\
 & Wilks & 99.9--100.0 \\
\bottomrule
\end{tabular}

\end{table}

\begin{figure}[!htbp]
\centering
\includegraphics[width=\textwidth]{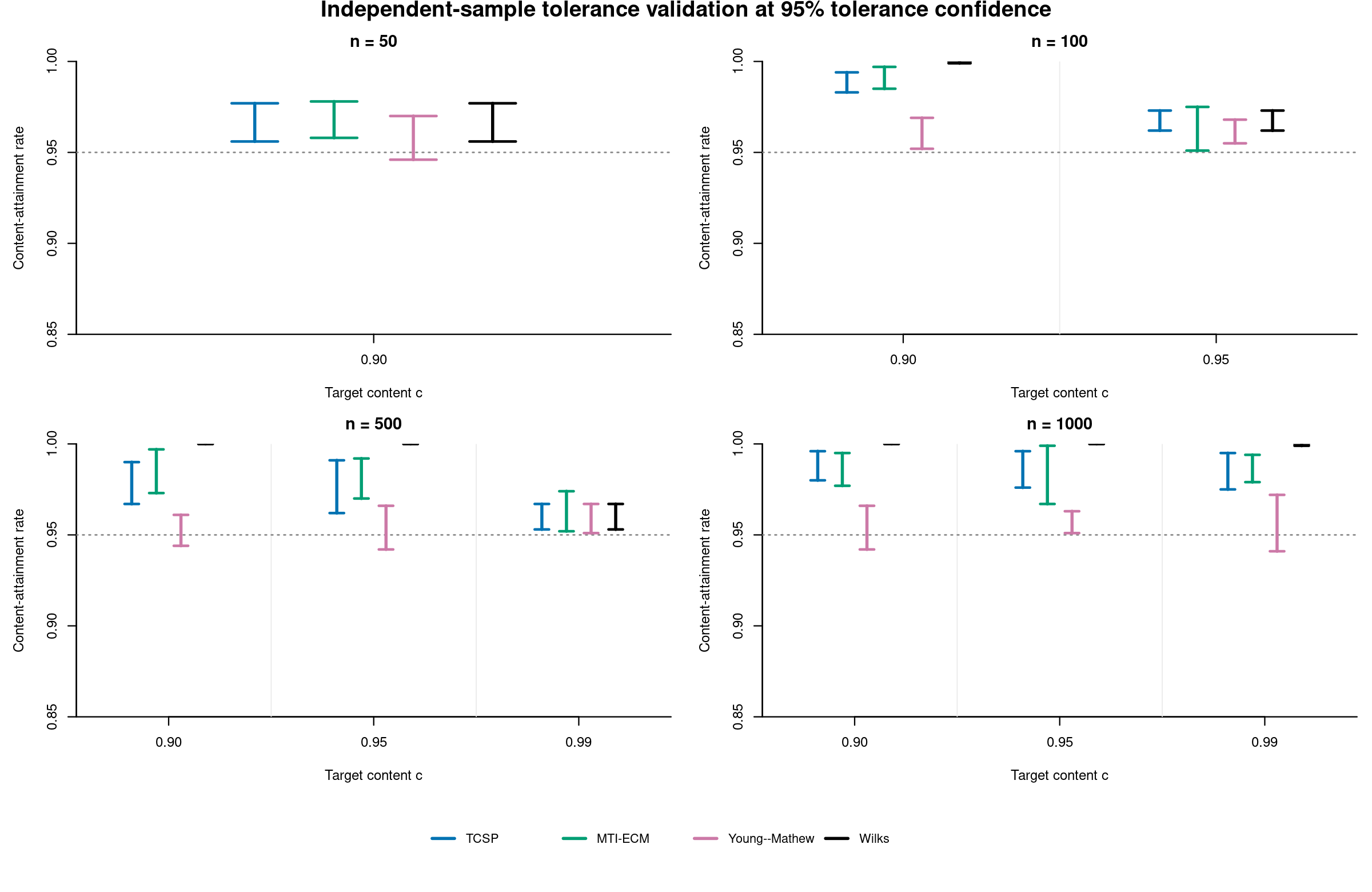}
\caption{\textbf{Content-attainment by sample size and target content.}
Intervals not produced are counted as nonattainment.
Vertical intervals in the content-attainment panels show the range across
distributional scenarios, and the axis is zoomed to \(0.85\)--\(1.00\).
Dashed reference lines mark nominal tolerance confidence \(0.95\).}
\label{fig:tolerance-validation-primary}
\end{figure}

\begin{figure}[!htbp]
\centering
\includegraphics[width=\textwidth]{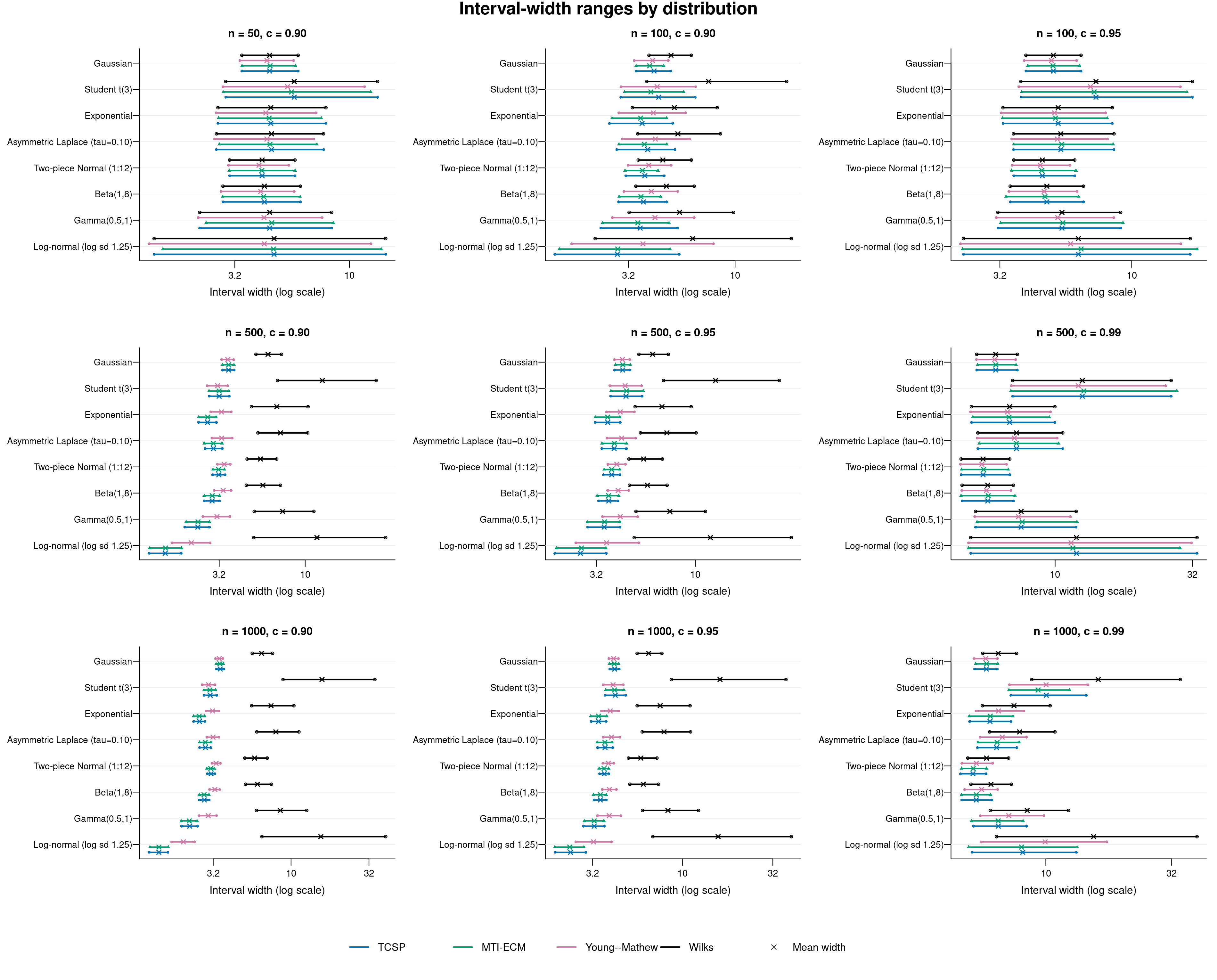}
\caption{\textbf{Interval-width ranges by distribution.} Horizontal intervals show empirical
2.5\%--97.5\% ranges of interval lengths within each distribution,
sample-size/content cell, and method. Crosses mark the mean interval width
across the 1000 paired replications. The horizontal axis is logarithmic to
display light-tailed, heavy-tailed, and strongly skewed cases in one figure;
the values remain raw interval lengths.}
\label{fig:tolerance-validation-width-ranges}
\end{figure}

In the completed run, TCSP content-attainment ranges from 95.3\% to 99.6\% across
distribution-level cells, compared with 95.1\% to 99.9\% for MTI-ECM,
94.1\% to 97.2\% for Young--Mathew, and 95.3\% to 100.0\% for Wilks.
Figures~\ref{fig:tolerance-validation-primary}
and~\ref{fig:tolerance-validation-width-ranges} show the same comparison without
pooling widths across distributional scenarios. The calibrated scan action
targets the requested finite-sample statement directly. In the independent
validation run, the selected adaptive MTI-ECM rule has observed content
attainment above the nominal tolerance confidence in every distribution-level
cell. MTI-ECM provides a less conservative
width-seeking comparison tied to the generalized-Bayes interval construction:
it also stays above nominal tolerance confidence in all 72 distribution-level
cells and has a smaller cellwise median width than TCSP in 38 cells. The closest
independent validation cells remain near the nominal boundary, so this evidence
should be read as repeated-sampling validation of the selected MTI-ECM rule
rather than as a distribution-free finite-sample certificate. Young--Mathew is
a strong nonparametric comparator and is slightly
shorter on the symmetric reference distributions, but it falls below nominal
tolerance confidence in nine skewed or heavy-tailed cells. In the six
asymmetric simulation distributions, TCSP and MTI-ECM are each shorter than
Young--Mathew in 36 of 54 cells. Wilks is simple and reliable, but it is often
much wider when the scan calibration does not force the full sample range.

The supplement gives the distribution-level content-attainment and width tables behind these
summaries, the selected MTI-ECM calibration summary, the TCSP retained counts,
and the full-range feasibility thresholds used to choose the simulation design.
The excluded small-sample cells fall outside the region where a two-sided
distribution-free tolerance statement can be certified at confidence \(0.95\).
Other variants are kept outside the primary comparison.

\section{Discussion}
\label{sec:discussion}

Probability content alone determines a class of interval placements. MPI selects
the mean-preserving member under the stated population conditions. MTI
generalizes MPI by making retained-mean balance a fixed coordinate:
\(\delta=0\) returns MPI, and admissible nonzero tilts specify other contiguous
content-\(c\) windows. Equal-tailed and shortest-contiguous intervals are
therefore distribution-specific members of the same placement family.

The tolerance contribution is a separate frequentist layer. A scan statistic
chooses an empirical retained count, after which the shortest closed
order-statistic window at that count is reported. This makes TCSP the formal
tolerance action and gives a class-specific minimum-width property within the
calibrated empirical scan class. The validation study shows that TCSP gives
reliable content attainment across the feasible design cells while avoiding much
of the width inflation of full-range Wilks intervals. The calibrated MTI-ECM
comparator links the empirical study to the MTI loss and can be shorter in many
cells, but its evidence is repeated-sampling validation of a selected rule
rather than a distribution-free finite-sample theorem.

The separation between target, computation, and certification is essential. MTI
identifies which fixed-content interval is being estimated. A generalized
posterior describes uncertainty for a fixed loss-defined target. A response
model, such as a direct Dirichlet-process posterior for \(F\), gives
model-based content probabilities. Tolerance confidence in TCSP comes from the
scan calibration. Moving a guarantee from one layer to another requires a
separate theorem or direct validation study.

Several limitations remain. The formal TCSP statement is iid, continuous, and
univariate, and it uses the closed order-statistic convention stated in the
paper. Boundary tilts can lead to one-sided limits or inactive roots, and tilts
outside the admissible range do not define a finite unrestricted target.
Regression-family and dynamic tolerance calibration, selected-action
asymptotics, exact scan recursion, and posterior-to-action transfer remain open
problems. These topics are better treated in a companion development focused on
endpoint regression and dynamic computation, where the evidence standards differ
from the finite-sample tolerance action studied here.

\clearpage
\bibliographystyle{plainnat}
\bibliography{refs}

\end{document}